\documentclass[superscriptaddress,pra,aps,twocolumn,10pt,longbibliography]{revtex4-1}
\usepackage{bm}
\usepackage{amsmath}
\usepackage{amssymb}
\usepackage{amsfonts}
\usepackage{graphicx}
\usepackage{grffile}
\usepackage{natbib}
\usepackage{epstopdf}
\usepackage[pdftex]{color}
\usepackage[dvipsnames]{xcolor}
\usepackage{hyperref}
\usepackage{verbatim}
\usepackage[normalem]{ulem}
\usepackage{float}
\usepackage[boxed]{algorithm}
\usepackage{algpseudocode}
\usepackage{soul}
\hypersetup{
    colorlinks=true,       % false: boxed links; true: colored links
    linkcolor=cyan,          % color of internal links
    citecolor=magenta,        % color of links to bibliography
    filecolor=magenta,      % color of file links
    urlcolor=cyan,           % color of external links
    runcolor=cyan
}

\newcommand{\beq}{\begin{eqnarray}}
\newcommand{\eeq}{\end{eqnarray}}
\newcommand{\bes} {\begin{subequations}}
\newcommand{\ees} {\end{subequations}}

\newcommand{\ignore}[1]{}

\graphicspath{{figs/}}

\begin{document}

\title{Characterizing local noise in QAOA circuits}

\author{Jeffrey Marshall}
\affiliation{QuAIL, NASA Ames Research Center, Moffett Field, California 94035, USA}
\affiliation{USRA Research Institute for Advanced Computer Science, Mountain View, California 94043, USA}

\author{Filip Wudarski}
\affiliation{QuAIL, NASA Ames Research Center, Moffett Field, California 94035, USA}
\affiliation{USRA Research Institute for Advanced Computer Science, Mountain View, California 94043, USA}

\author{Stuart Hadfield}
\affiliation{QuAIL, NASA Ames Research Center, Moffett Field, California 94035, USA}
\affiliation{USRA Research Institute for Advanced Computer Science, Mountain View, California 94043, USA}

\author{Tad Hogg}
\affiliation{QuAIL, NASA Ames Research Center, Moffett Field, California 94035, USA}
\affiliation{USRA Research Institute for Advanced Computer Science, Mountain View, California 94043, USA}

\begin{abstract}
Recently Xue et al. [arXiv:1909.02196 \cite{qaoa-noise}] demonstrated numerically that QAOA performance varies as a power law in the amount of noise under certain physical noise models. In this short note, we provide a deeper analysis of the origin of this behavior. In particular, we provide an approximate closed form equation for the fidelity and cost in terms of the noise rate, system size, and circuit depth. As an application, we show these equations accurately model the trade off between larger circuits which attain better cost values, at the expense of greater degradation due to noise.
\end{abstract}
\maketitle

\section{Introduction}

We study noise in Quantum Approximate Optimization Algorithm (QAOA) circuits \cite{qaoa}.
We will consider the original formulation of QAOA, with a transverse field mixer, though a generalization to the Quantum Alternating Operator Ansatz exists \cite{qaoa-nasa}. In the version we study, a QAOA circuit of $N$ qubits is specified by a cost Hamiltonian diagonal in the computational basis \cite{note-problems}

\begin{equation}
    H_c = \sum_{i=1}^N h_i \sigma_i^z + \sum_{j>i=1}^N J_{ij}\sigma_i^z \sigma_j^z +\dots,
    \label{eq:cost}
\end{equation}
as well as  $2d$ angles ($\vec{\gamma},\vec{\beta}$), $d$ being the high-level depth of the circuit $U$ (i.e., number of QAOA rounds):
\begin{equation}
    U(\vec{\gamma},\vec{\beta}):= \prod_{k=1}^d e^{-i \beta_k H_x}e^{-i \gamma_k H_c}.
\end{equation}
The Hamiltonian $H_x$ with off-diagonal elements, known as the `mixing' Hamiltonian, is given by $ H_x = \sum_{i=1}^N \sigma_i^x$.
We will assume throughout (though this is not necessary) the initial state is an equal superposition over all computational basis states, $|\psi_0\rangle=|+\rangle^{\otimes N}$, with $|+\rangle=\frac{1}{\sqrt{2}}(|0\rangle + |1\rangle)$.

The goal of QAOA is to find approximate solutions to optimization problems phrased as Eq.~\eqref{eq:cost}, which can be achieved by finding `good' angles $(\vec{\gamma},\vec{\beta})$ in order to minimize or maximize the expected cost,
\begin{equation}
    C(\vec{\gamma},\vec{\beta}) = \langle \psi_0|U^\dag H_c U|\psi_0\rangle.
    \label{eq:ideal-cost}
\end{equation}
A very natural question to ask is how noise affects the performance of a QAOA algorithm. There are of course several possible ways to address this, from misspecification of angles, state preparation and measurement errors, environmental noise etc. 
Recently Xue et al. \cite{qaoa-noise} introduced a model for how local environmental noise (acting on each qubit separately and independently) affects the circuit output which is amenable to a mathematical analysis, by decomposing the density matrix as a sum of pure states.

Following this work, we will take a simple, yet realistic model to study the behaviour of the cost expectation function Eq.~\eqref{eq:ideal-cost} when a layer of local depolarizing noise acts after each QAOA round.
Xue et al. demonstrated that under such a model, though gradients become flatter, the general landscape does not change much \footnote{We also find the same general effect in our simulations: optimal angles in the noiseless case are approximately optimal angles when performing optimization with noise. We use the BFGS algorithm to optimize QAOA angles, using random initial angles each time. We also repeat over several initializations to help avoid being trapped in local optima. Note that optimized angles for different circuit depths in general are completely different, since we optimize over the entire set of angles, and not round by round. In our simulations the angles are in range $[0,2\pi]$.}. This simplification allows us to answer in more detail questions pertaining to how local noise causes deviations in the expected cost and what the relative trade-offs are with respect to system size, circuit size (depth) and the noise rate.

\section{Noise model}
\begin{figure}
    \centering
    \includegraphics[width=0.46\textwidth]{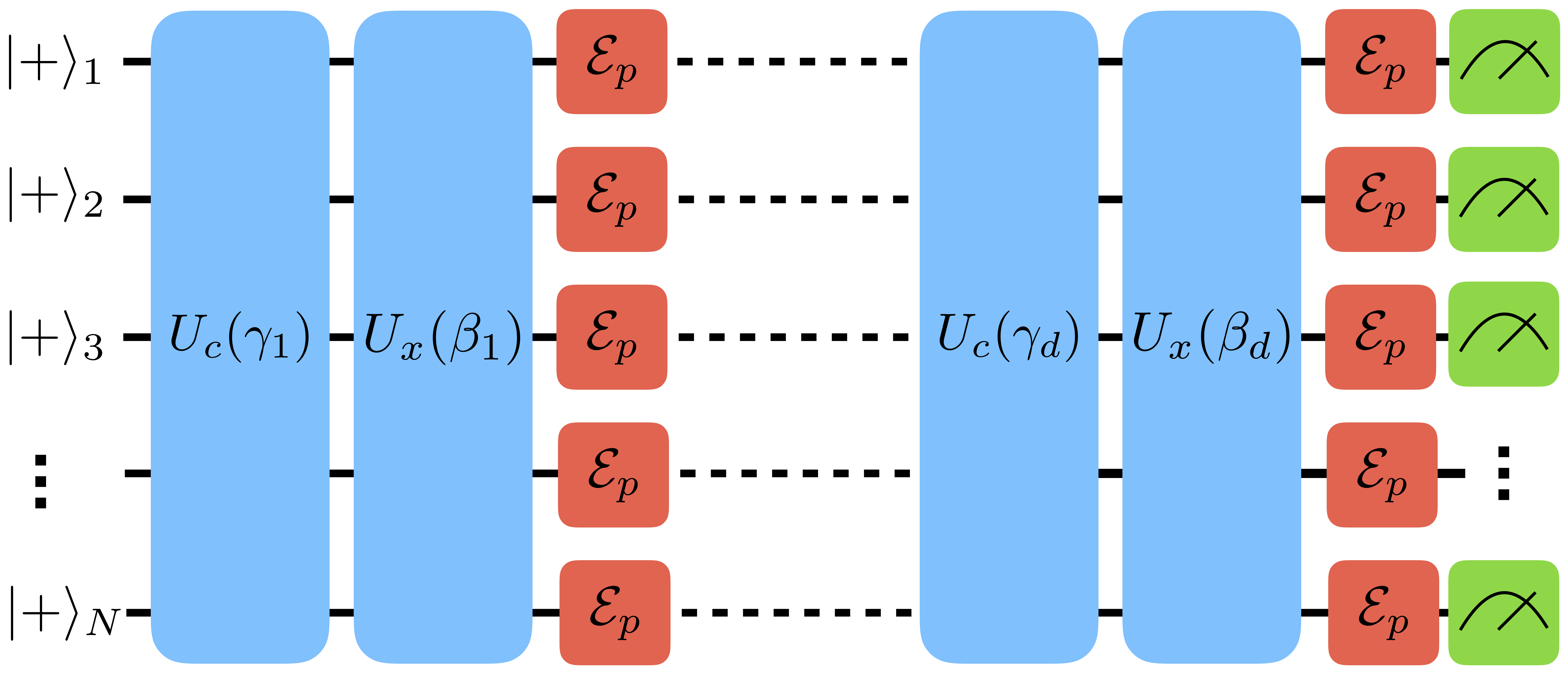}
    \caption{Noisy QAOA architecture used in simulations. Each of the $d$ blocks 
    of QAOA is composed of $U_c(\gamma_k) = e^{-i\gamma_k H_c}$ and $U_x(\beta_k) = e^{-i\beta_k H_x}$ followed by noise channel $\mathcal{E}_p$ that acts locally on each qubit with the same ``noise'' strength $p$.}
    \label{fig:qaoa}
\end{figure}

Similar to Ref.~\cite{qaoa-noise}, we assume that after each QAOA round a layer of local noise $\mathcal{E}_p$ is applied to each qubit (see Fig.~\ref{fig:qaoa}), of the form
\begin{equation}
    \label{eq:kraus}
    \mathcal{E}_p\rho = (1-p)\rho +\frac{p}{M} \sum_{j=1}^{M} K_j \rho K_j^\dag,
\end{equation}
where $p\in [0,1]$ 
\footnote{For now we assume no restriction on the $K_j$ other than that $\mathcal{E}_p$ is a quantum map, and so to preserve the trace, $ \sum_j K_j^\dag K_j = M\mathbb{I}$.}. 
With this freedom of $p$, any quantum map can be phrased as in Eq.~(\ref{eq:kraus}), but we will focus on cases with $p<1$ where there is a term proportional to the  state $\rho$, which includes depolarizing and dephasing noise (among other physically relevant noise models) 
\footnote{For qubit-local depolarizing noise, $M=4$, with $K_i=\sigma_i$, the identity and three Pauli $x,y,z$ operators. Although we could move the identity term out of the sum in Eq.~\eqref{eq:kraus} (and changing `$1-p$' to `$1-3p/4$'), it is convenient to leave it in as the second term is simply the maximally mixed state $\mathbb{I}$, which is i) easier to handle mathematically, and ii) provides the natural interpretation of the channel, that one either receives the ideal state with probability $(1-p)$, or the maximally mixed state.}.

To generalize slightly to the case of $N$ qubits, we write $\mathcal{E}_p^{(n)}$ (for $n=1,\dots ,N$) to denote the noise channel (of form Eq.~\eqref{eq:kraus}) acting on qubit $n$, and the identity channel on all of the other $N-1$ qubits \footnote{To be more formal, we can write $\mathcal{E}_p^{(n)}=\mathrm{Id}\otimes \dots \otimes \mathrm{Id} \otimes \mathcal{E}_p \otimes \mathrm{Id}\otimes \dots \otimes \mathrm{Id}$, with the map $\mathcal{E}_p$ in the $n$-th position, and the remaining $N-1$ identity $\mathrm{Id}$ channels acting on the other qubits, $\mathrm{Id}X=X$.}. In this way, a single round of QAOA followed by the noise layer is of the form
\begin{equation}
    \rho_1=\prod_{n=1}^N \mathcal{E}_p^{(n)} \left[ U(\gamma,\beta)\rho_0 U(\gamma,\beta)^\dag\right],
    \label{eq:rho1}
\end{equation}
where we just specify one set of angles $\gamma,\beta$ (i.e. $d=1$), and writing $\rho_0=|\psi_0\rangle \langle \psi_0|$. Note that the noise layer can also be written as $\prod_{n=1}^N \mathcal{E}_p^{(n)} = \bigotimes_{n=1}^N \mathcal{E}_p$. At this level, we can interpret the probability of the noise not acting at all as simply $(1-p)^N$, or, if one repeats this for $d$ rounds, it is  $(1-p)^{dN}$.

We focus initially on $d=1$. Let us define $|\psi_1\rangle = U|\psi_0\rangle$ (i.e. the ideal, noiseless QAOA-1 output). We can interpret the map $\mathcal{E}_p^{(n)}|\psi_1\rangle\langle \psi_1 |$ statistically, as in Ref.~\cite{qaoa-noise}, as applying noise operator $K_j^{(n)}$ (defined as identity on all qubits, and $K_j$ on $n$-th qubit \footnote{$K_j^{(n)}=\mathbb{I}\otimes \dots \otimes \mathbb{I} \otimes K_j \otimes\mathbb{I}\otimes \mathbb{I}$, with $K_j$ on the $n$-th qubit.}) 
with probability $p_j^{(n)}=\frac{p}{M}\|K_j^{(n)}|\psi_1\rangle \|^2$, resulting in the quantum state $|\psi_j^{(n)}\rangle=K_j^{(n)}|\psi_1\rangle/\|K_j^{(n)}|\psi_1\rangle \|$. For ease of notation (though this restriction is not necessary), and considering we will later focus on such models, let us assume the $K_j$ are unitary (e.g. as in depolarizing noise), in which case we can drop the normalization coefficients, so that $p_j^{(n)}=\frac{p}{M}$. Let us also adopt the notation $|\psi_{\vec{j}}^{(\vec{n})}\rangle=|\psi_{j_1, j_2,\dots}^{(n_1,n_2,\dots)}\rangle = K_{j_1}^{(n_1)}K_{j_2}^{(n_2)}\dots |\psi_1\rangle$, where the vector notation for $\vec{j},\vec{n}$ should be clear.

Then Eq.~\eqref{eq:rho1} can be written as
\begin{equation}
    \rho_1=\sum_{m=0}^N (1-p)^{N-m}\left(\frac{p}{M}\right)^m \sum_{\vec{j}_m,\vec{n}_m}|\psi_{\vec{j}_m}^{(\vec{n}_m)}\rangle \langle \psi_{\vec{j}_m}^{(\vec{n}_m)}|,
    \label{eq:rho1_sum}
\end{equation}
where $\vec{n}_m,\vec{j}_m$ are length $m$ vectors, with each entry in $\vec{n}_m$ distinct (i.e. no repeats). That is, $m$ specifies the total number of noise operators acting; $m=0$ is the case where no noise acts, and the other extreme, where $m=N$ means every qubit has a noise operator applied.
The second sum in Eq.~\eqref{eq:rho1_sum} is therefore over $M^m \times {N \choose m}$ unique terms, using that each index in $\vec{j}$ can run from 1 to $M$ (repeats allowed), and that $\vec{n}$ has elements ranging from 1 to $N$, but with the restriction each element is unique.

\section{Approximate analytical expressions for QAOA performance under noise}

A natural question is how $\rho_1$ compares to the ideal QAOA-1 output, $|\psi_1\rangle$. Two key quantities of interest are the fidelity between the two, and the difference in expected cost. 

\subsection{Fidelity}
First, let us consider the fidelity, or overlap, between $\rho_1$ and $|\psi_1\rangle$
\begin{equation}
    F_1 = \langle \psi_1 | \rho_1 |\psi_1\rangle.
    \label{eq:fidelity}
\end{equation}
In Ref.~\cite{qaoa-noise}, it is posited, and shown numerically for the parameters studied, that the fidelity fits to a generic polynomial $F_1=(1-p)^{\delta N}$ 
\footnote{In Ref.~\cite{qaoa-noise}, the `$N$' in $F_1$ is the number of noise operations applied (number of gates). In our model, for QAOA-1, $N$ is indeed the number of qubits, since we apply noise after each round and individually on each qubit. }.
One aim of this note is to justify such an equation in the case of local unitary noise, such as depolarizing/dephasing noise. 

First we write, using Eq.~\eqref{eq:rho1_sum} and \eqref{eq:fidelity},
\begin{equation}
  \sum_{\vec{j}_m,\vec{n}_m} |\langle \psi_1 |\psi_{\vec{j}_m}^{(\vec{n}_m)}\rangle|^2  = M^m {N\choose m} \overline{|\langle \psi_1 |\psi_{\vec{j}_m}^{(\vec{n}_m)}\rangle|^2},
\end{equation}
where the `overline' represents the average (mean) value of the overlaps (i.e. over all possible $\vec{n}_m, \vec{j}_m$).

\begin{figure}
    \centering
    \includegraphics[width=0.98\columnwidth]{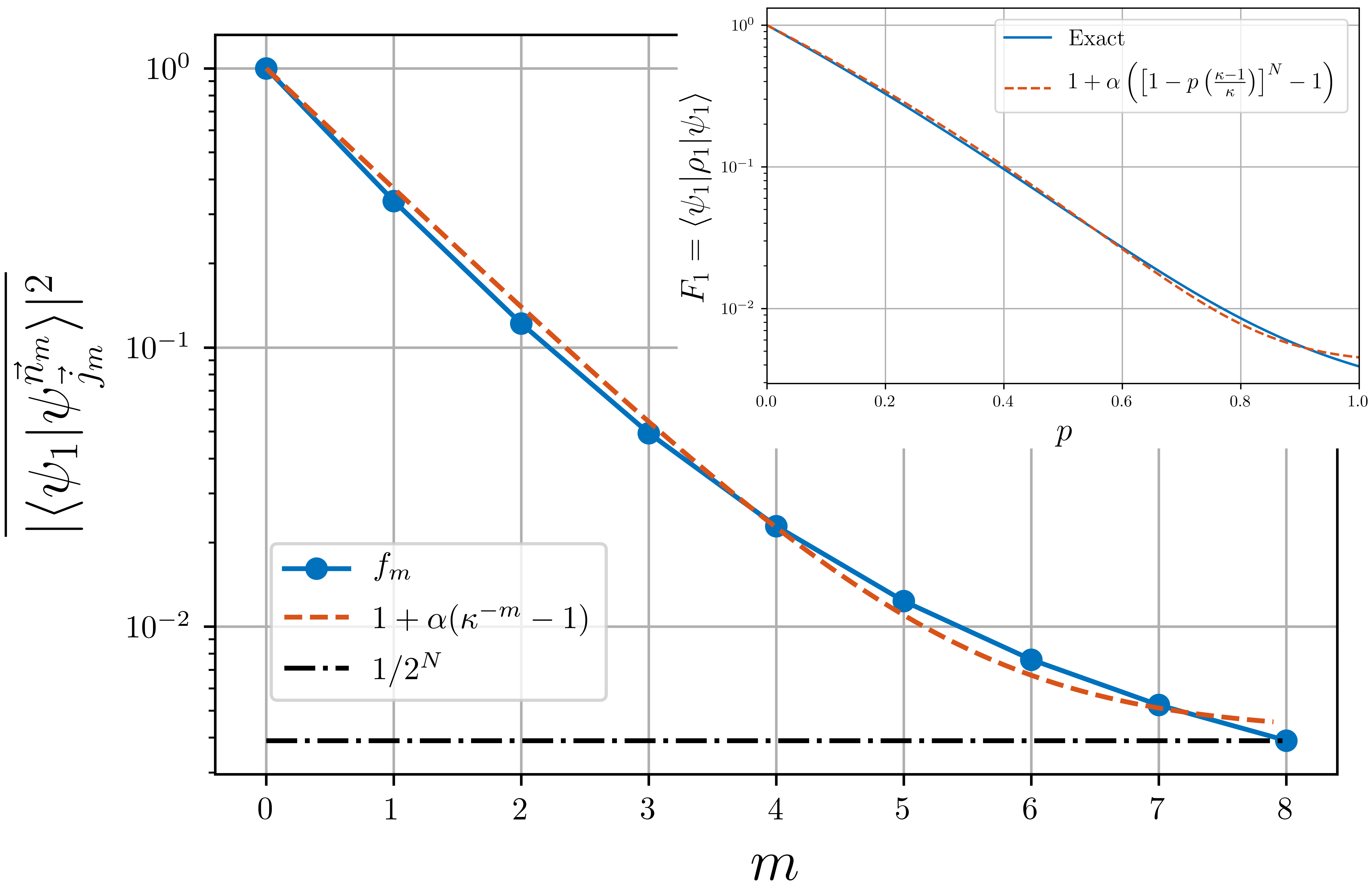}
    \caption{Plot of the average value of overlaps $|\langle \psi_1 | \psi_{\vec{j}_n}^{(\vec{n}_j)} \rangle |^2$ for a typical $N=8$ problem instance \cite{note-problems} under depolarizing noise.
    Solid blue is through the exact mean overlap, $f_m$, and red dash is fit $\kappa^{-m}$ to this using non-linear least-squares (on the log of the data). 
     The horizontal black dash-dot line is the Haar random overlap value. The fit gives $\alpha=0.9958;\kappa=2.71$. \textbf{Inset}: We plot for the depolarizing channel, the exact (blue) fidelity (Eq.~\eqref{eq:fidelity}) as a function of $p$. We see the fitted curve (red dash) Eq.~\eqref{eq:fidelity_binomial}, with parameters extracted from $f_m$ in the main figure, matches very well for nearly all $p$. }
    \label{fig:overlaps_m}
\end{figure}

A reasonable assumption is that the average on the right depends only on the parameter $m$, and can therefore be written as a function $f_m=\overline{|\langle \psi_1 |\psi_{\vec{j}_m}^{(\vec{n}_m)}\rangle|^2}$.
With this, we can write the fidelity Eq.~\eqref{eq:fidelity} as
\begin{equation}
    F_1 = \sum_{m=0}^N {N\choose m} (1-p)^{N-m} p^m f_m.  
\end{equation}
An example of this is shown in Fig.~\ref{fig:overlaps_m} for a typical problem instance.

It now becomes clearer where the binomial form discovered in Ref.~\cite{qaoa-noise} comes from. First note we have $f_0=1$ by definition, and we expect $f_N \approx 2^{-N}$ (the overlap squared of two random states). While it is tempting to set here $f_m=2^{-m}$, 
as the scaling is in general noise dependent, we use the slightly less restrictive form (satisfying $f_0=1$) and write instead $f_m\approx 1+\alpha(\kappa^{-m}-1)$, and so, by the binomial theorem
\begin{equation}
    F_1 \approx 1+\alpha \left(\left[ 1-p\left( \frac{\kappa -1}{\kappa}\right)\right]^N - 1\right).
    \label{eq:fidelity_binomial}
\end{equation}
Due to the generality of the form $f_m$ we expect that many systems of interest will follow Eq.~\eqref{eq:fidelity_binomial}, though we mention that in our simulations here, we only consider depolarizing noise.

Notice that this equation, derived from a few reasonable approximations, predicts a slightly different dependence on $N$ compared to the assumed form from Ref.~\cite{qaoa-noise}, though both are equivalent for sufficiently small noise rate $p$.
Indeed, in the limit of small $p\ll 1$, as may be expected in a reasonable quantum circuit, one can relate the exponent $\delta$ found in Ref.~\cite{qaoa-noise} as $\delta=\alpha \frac{\kappa - 1}{\kappa}$. If we take, as in Fig.~\ref{fig:overlaps_m}, $\alpha=0.996, \kappa=2.71$, then $\delta=0.63$. In this case, we nearly have $f_m=\kappa^{-m}$, and simply $F_1=[1-p(\frac{\kappa-1}{\kappa})]^N$.

As demonstrated in the inset of Fig.~\ref{fig:overlaps_m}, our formula (Eq.~\eqref{eq:fidelity_binomial}) also applies to when the noise is dominant (e.g. $p>0.5$), and not just for `small' $p$. 
Notice that the full curve over $p\in[0,1]$ cannot be replicated by a formula $(1-p)^{\delta N}$, which does not give the appropriate behaviour in the large $p$ limit (it would diverge to $-\infty$ if plotted on the same scale as the inset). 

\subsection{Cost}
The expected cost for our noisy QAOA-1 is given by $\langle H_c\rangle = \mathrm{Tr}[H_c \rho_1]$. For well chosen angles ($\gamma,\beta$) using QAOA to minimize, we expect the cost to increase under the noise, i.e. $C_{\mathrm{noise}}^{(1)}=\mathrm{Tr}[H_c \rho_1]>\langle \psi_1 | H_c|\psi_1\rangle=C_{\mathrm{ideal}}^{(1)}$. Ref.~\cite{qaoa-noise} posits that the dependence can be written as 
\begin{equation}
    C_{\mathrm{noise}}^{(1)}=(1-p)^{ \eta N}C_{\mathrm{ideal}}^{(1)},
\end{equation}
for some $\eta>0$, where we use, matching the condition in the large noise limit $p\rightarrow 1$, the Haar random expectation $C_{\mathrm{Haar}}=\frac{1}{2^N}\mathrm{Tr}[H_c]=0$ by Eq.~\eqref{eq:cost} \footnote{In Ref.~\cite{qaoa-noise}, if there are terms proportional to the identity $a\mathbf{I}$ in $H_c$, so that $\mathrm{Tr}[H_c]\neq 0$, the effect is to add a term $(1-(1-p)^{\eta N})a$ to $C_{\mathrm{noise}}$. In our analysis we do not need to do this as the equation we derive handles such a case through the freedom of parameters}.
For well chosen angles, one expects $C_{\mathrm{ideal}}^{(1)} < 0$.

We perform a similar analysis as above via Eq.~\eqref{eq:rho1_sum}, and consider the terms with the same number of noise operators acting
\begin{equation}
   \sum_{\vec{j}_m,\vec{n}_m}\langle \psi_{\vec{j}_m}^{(\vec{n}_m)}| H_c | \psi_{\vec{j}_m}^{(\vec{n}_m)}\rangle = M^m{N\choose m} \overline{\langle \psi_{\vec{j}_m}^{(\vec{n}_m)}| H_c | \psi_{\vec{j}_m}^{(\vec{n}_m)}\rangle}.
\end{equation}

For $m=0$, the average term on the right is precisely $C_{\mathrm{ideal}}$, and for $m=N$ it can be approximated by the Haar random expectation, which here is 0.
As before, if this quantity on the right only depends (to a good enough approximation) on $m$, we can replace it by a function $c_m=\overline{\langle \psi_{\vec{j}_m}^{(\vec{n}_m)}| H_c | \psi_{\vec{j}_m}^{(\vec{n}_m)}\rangle}$.
This structure suggests writing, as before,
$c_m \approx  \alpha + \tilde{\alpha} \chi^{-m}$,
where $\alpha+\tilde{\alpha} \approx C_{\mathrm{ideal}}^{(1)}$. 

\begin{figure}
    \centering
    \includegraphics[width=0.98\columnwidth]{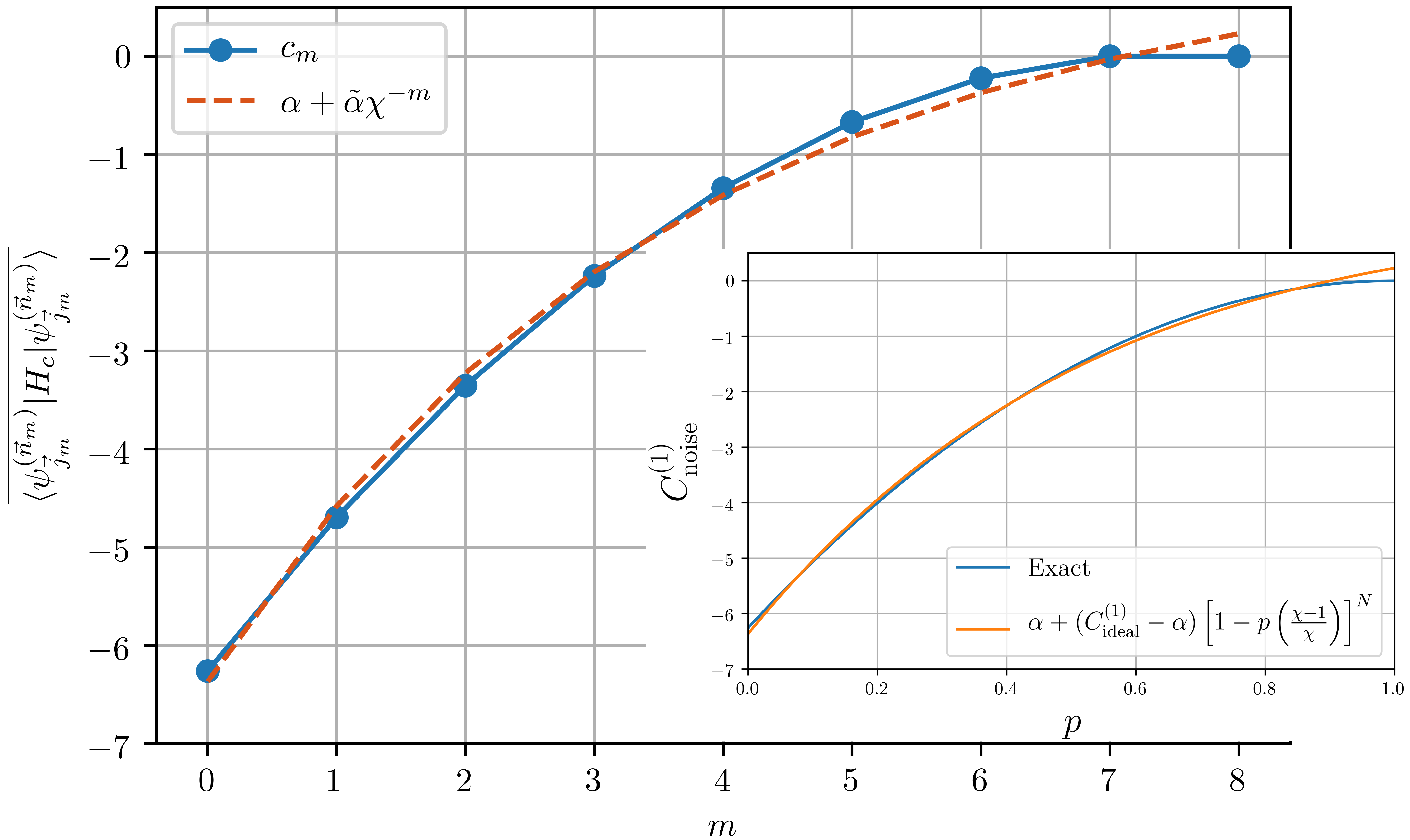}
    \caption{Plot of the average cost expectation values $\langle \psi_{\vec{j}_m}^{(\vec{n}_m)}| H_c | \psi_{\vec{j}_m}^{(\vec{n}_m)}\rangle$ for a typical $N=8$ problem instance \cite{note-problems} under depolarizing noise.  Blue (solid) line is through the exact $c_m$, and red (dash) is fit $\chi^{-m}$ to this using non-linear least-squares. The fit gives $\alpha=1.04;\tilde{\alpha}=-7.41;\chi=1.32$. \textbf{Inset}: We plot for the depolarizing channel, the exact (blue-solid) cost as a function of $p$. We see the fitted curve (red-dash) Eq.~\eqref{eq:cost_binomial}, with parameters extracted from $c_m$ in the main figure, matches very well for nearly all $p$. Note, the ground-eigenenergy of $H_c$ is -14.}
    \label{fig:costs_m}
\end{figure}
With this, again by the binomial theorem,
\begin{equation}
    C_{\mathrm{noise}}^{(1)}\approx \alpha + (C_{\mathrm{ideal}}^{(1)}-\alpha) \left[1 - p\left(\frac{\chi - 1}{\chi} \right)\right]^N,
    \label{eq:cost_binomial}
\end{equation}
showing a slightly different form from Ref~\cite{qaoa-noise}.
In the small $p$ limit however, one can extract the relation $\eta = \frac{C_{\mathrm{ideal}}^{(1)}-\alpha}{C_{\mathrm{ideal}}^{(1)}}\frac{\chi-1}{\chi}$. Using the values from Fig.~\ref{fig:costs_m}, we get $\eta=0.28$ for this example.

As in the case of fidelity, our equation also works to good accuracy in nearly the full range $p\in[0,1]$, i.e. in particular where $p>0.5$ is not a `small' parameter, which is shown in the inset of Fig.~\ref{fig:costs_m}.

\subsection{QAOA-$d$}
For QAOA with $d$ rounds (QAOA-$d$) we can extend the above analysis, which sheds light onto the depth vs.~noise trade-off. Again, the model we follow is a round of QAOA, followed by $N$ local noise channels, repeated $d$ times, as in Fig.~\ref{fig:qaoa}. Each noise channel is assumed to act identically on each qubit, and in each round.
This means we can write an exact expression for $\rho_d$ (i.e. in the form Eq.~\eqref{eq:rho1_sum}), where one replaces $N$ with $Nd$, since this is the total possible number of noise terms (the $K_j^{(n)}$) which can act:
\begin{equation}
    \rho_d = \sum_{m=0}^{Nd}(1-p)^{Nd-m}\left(\frac{p}{M}\right)^m\sum_{\vec{j}_m,\vec{l}_m,\vec{n}_m} |\psi_{\vec{j}_m, \vec{l}_m}^{(\vec{n}_m)}\rangle \langle \psi_{\vec{j}_m, \vec{l}_m}^{(\vec{n}_m)} |,
    \label{eq:rho_d}
\end{equation}
where the second sum has a new index, $\vec{l}$ representing the QAOA layer (round) in which the noise operator acts. That is, the elements of the length $m$ vectors $(\vec{l}_m,\vec{n}_m,\vec{j}_m)$ tell us respectively which layer, qubit and operator the noise is acting.
The second sum is over $M^m {Nd \choose m}$ terms. For notational transparency, the general form of the term in the second sum is:
\begin{equation}
    |\psi_{\vec{j}_m, \vec{l}_m}^{(\vec{n}_m)}\rangle = K^{\vec{n}}_{\vec{j},d} U_d\dots K^{\vec{n}}_{\vec{j},2}  U_2 K^{\vec{n}}_{\vec{j},1}U_1|\psi_0\rangle,
\end{equation}
where $U_k$ is the ideal QAOA unitary in round $k$, and $K_{\vec{j},k}^{\vec{n}}=K_{j_{m_k}}^{n_{m_k}}\dots K_{j_{m_1}}^{n_{m_1}}$ are the noise operators acting in that round (with $\sum_k m_k = m$).
Following the form of Eqs.~\eqref{eq:fidelity_binomial}, \eqref{eq:cost_binomial}, one would simply replace `$N$' by `$Nd$' for $F_d$ and $ C_{\mathrm{noise}}^{(d)}$. Note that the fitting parameters e.g. $\alpha,\chi$, also depend on the depth $d$, since different depths correspond to completely different circuits.

\begin{figure}
    \centering
    \includegraphics[width=0.98\columnwidth]{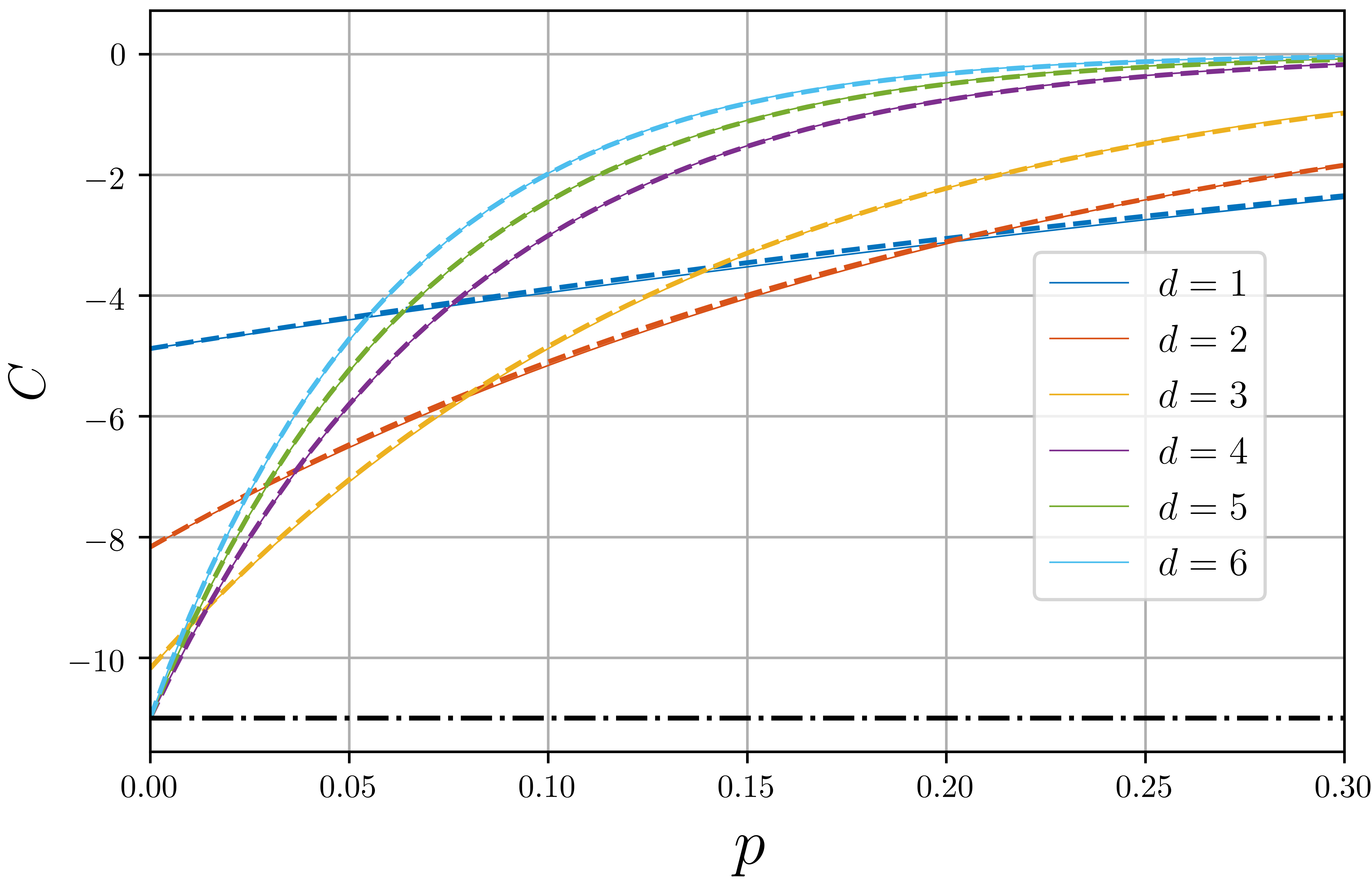}
    \caption{Cost $C=C_{\mathrm{noise}}^{(d)}(p)$ as a function of depolarizing probability for various circuit depths $d$, for a typical $N=6$ instance \cite{note-problems}. For larger values of $p>0.3$ the curves do not cross again, and all converge to the Haar random cost of 0 as $p\rightarrow 1$. The black dash-dot line is the optimal cost (ground state), which is found to a good approximation for $d\ge 4$ in the noiseless case ($p=0$). Each colored dash line is from a non-linear least squares fit to a function of form $C_{\mathrm{noise}}^{(d)}$ (i.e. Eq.~\eqref{eq:cost_binomial} with $N\rightarrow Nd$), and matches the data (colored solid lines) very well.}
    \label{fig:C_pd}
\end{figure}

Of practical interest is the relative trade-off between performing a greater number of rounds of QAOA, thus ideally obtaining a lower cost, but at the expense that there is additional noise acting to raise the cost. Clearly for large $p$, there is no benefit to going beyond $d=1$ since the output will be close to the maximally mixed state. In Fig.~\ref{fig:C_pd} we see the complex interplay between depth $d$ and noise level $p$. Only for very small noise levels is it beneficial to go to large depths. Once the depolarizing probability exceeds 2\% ($p=0.02$) there is no reason to go beyond $d=3$. For this example, if $p>0.25$, the curves have inverted from the order at $p=0$, i.e., here $d=1$ is always the best choice.

If one can estimate the parameters $(\alpha,\chi)$ for a few circuit depths of interest (e.g. as is done in Fig. \ref{fig:costs_m}), it is possible therefore to determine optimal depth circuit for a given noise rate $p$.

\section{Conclusion}
We have extended a previous work on noise in QAOA. Ref.~\cite{qaoa-noise} analyzed noise in QAOA circuits by decomposing the density operator as a sum of pure states with different numbers of noise operators acting. We follow this, and through approximations backed up by numerical simulation, we show that a slightly different binomial form is achieved, thus generalizing the posited fidelity and cost functions found in that previous work. We find this equation predicts a trade-off between noise level and depth, i.e., accuracy of optimization, showing that for large enough noise rates, it is best to keep a shorter circuit. 

This line of research opens up the route for further studies into noise in optimization circuits. In particular, the proposed approach can be used as one of many benchmarks in a suite of easily implementable algorithms tailored to assess performance of quantum devices. Since the method is scalable and tunable with respect to noise parameters - depth and number of qubits can play a proxy role for noise strength - one may compare theoretical predictions (subjected to a reasonable noise model) with experimental outcomes.

\acknowledgments
We are grateful for support from NASA Ames Research Center, the AFRL Information Directorate under grant F4HBKC4162G001, the Office of the Director of National Intelligence (ODNI) and the Intelligence Advanced Research Projects Activity (IARPA), via IAA 145483, and NASA Academic Mission Services, Contract No. NNA16BD14C.
We used QuTiP in our simulations \cite{qutip1}.
The views and conclusions contained herein are those of the authors and should not be interpreted as necessarily representing the official policies or endorsements, either expressed or implied, of ODNI, IARPA, AFRL, or the U.S. Government. The U.S. Government is authorized to reproduce and distribute reprints for Governmental purpose notwithstanding any copyright annotation thereon.

\bibliography{refs.bib}

\begin{thebibliography}{12}
\bibitem [{not()}]{note-problems}%
  \BibitemOpen
  \href@noop {} {}\bibinfo {note} {{The `dots' in Eq.~\eqref{eq:cost} allow in
  general higher order terms in the $\sigma^z$, though in our simulations we
  only consider up to quadratic terms. In our simulations we use fully
  connected problems of various sizes, without local fields $(h_i=0)$, where
  $J_{ij}\in\{-1,1\}$ randomly chosen}}\BibitemShut {NoStop}%
\end{thebibliography}
\end{document}